\documentclass[12pt]{article}
\usepackage{putex}
\usepackage{graphicx}

\begin{document}

\preprint{PUPT-2284}

\institution{PU}{Joseph Henry Laboratories, Princeton University, Princeton, NJ 08544}

\title{Low-temperature behavior of the Abelian Higgs model in anti-de Sitter space}

\authors{Steven S. Gubser and Abhinav Nellore}

\abstract{We explore the low-temperature behavior of the Abelian Higgs model in $AdS_4$, away from the probe limit in which back-reaction of matter fields on the metric can be neglected.  Over a significant range of charges for the complex scalar, we observe a second order phase transition at finite temperature.  The symmetry-breaking states are superconducting black holes.  At least when the charge of the scalar is not too small, we observe at low temperatures the emergence of a domain wall structure characterized by a definite index of refraction.  We also compute the conductivity as a function of frequency.}

\date{October 2008}

\maketitle

\tableofcontents

\section{Introduction and summary}
\label{INTRODUCTION}

The Abelian Higgs model in four-dimensional anti-de Sitter space ($AdS_4$) is specified by the action
 \eqn{Lagrangian}{
  S = {1 \over 2\kappa^2} \int d^4 x \, \sqrt{-g}\left[ R - 
    {1 \over 4} F_{\mu\nu}^2 - |(\partial_\mu - i q A_\mu) \psi|^2 - 
    V(|\psi|) \right] \,,
 }
where $V$ depends only on the magnitude of the complex scalar field $\psi$, not its phase.  Among the simplest solutions to the classical equations following from the action \eno{Lagrangian} is the $AdS_4$-Reissner-Nordstrom black hole solution, hereafter RNAdS:
 \eqn{RNAdS}{
  ds^2 &= e^{2A} \left( -h dt^2 + (dx^1)^2 + (dx^2)^2 \right) + 
    {dr^2 \over h}  \cr
  A &= {r \over L} \qquad 
  h = 1 - \epsilon L \kappa^2 e^{-3r/L} + 
    {\rho^2 \kappa^4 \over 4} e^{-4r/L}  \cr
  \Phi &= \rho \kappa^2 \left( e^{-r_H/L} - e^{-r/L} \right) 
   \qquad \psi = 0 \,.
 }

In \cite{Gubser:2008px}, following earlier work \cite{Gubser:2005ih}, it was suggested that there are black hole solutions to the classical equations of motion following from \eno{Lagrangian} which spontaneously break the $U(1)$ gauge symmetry associated with phase rotations of $\psi$.  The symmetry-breaking solutions can be thought of as superconducting black holes.  More precisely, there is a superconducting condensate of the scalar field $\psi$ which floats above the horizon.  A calculation of the critical temperature $T_c$ below which black hole superconductivity occurs was outlined in \cite{Gubser:2005ih} and carried through for a few choices of parameters such as $q$ and the anti-de Sitter space radius $L$.  The conformal field theory (CFT) dual to a superconducting black hole of this type is a symmetry-breaking phase whose order parameter is $\langle {\cal O}_\psi \rangle$, where ${\cal O}_\psi$ is the operator dual to $\psi$.  In the simplest setup, the phase rotations of ${\cal O}_\psi$ are a global symmetry of the boundary theory; however, one can weakly gauge this $U(1)$ symmetry, and then the symmetry-breaking in the boundary theory can also be described as superconductivity.  This transition from breaking a global symmetry to breaking a gauge symmetry has a precise parallel in BCS theory: all the dynamics of electrons and phonons can be carried out in ignorance of the gauge interactions; but once the condensate Cooper pairs are formed, one can go back and note that it breaks the abelian gauge symmetry.  As has been emphasized in \cite{Weinberg:1986cq}, many---though not all---of the macroscopic features of superconductors can be understood as a consequence of an effective field theory treatment of the breaking of the abelian gauge symmetry.

In \cite{Hartnoll:2008vx}, the picture of superconducting black holes was fleshed out in the limit of large $q$, where the matter fields do not back-react on the geometry.  It was further shown in \cite{Hartnoll:2008vx} that for the choice
 \eqn{VQuadratic}{
  V = -{6 \over L^2} + m^2 |\psi|^2 \,,
 }
with $m^2 L^2 = -2$, the conductivity defined from a two-point function of the conserved current dual to $A_\mu$ exhibits gap-like behavior.  This feature, remarkable for its qualitative (and perhaps semi-quantitative) similarity to real-world superconductors, cannot be anticipated on general grounds of symmetry-breaking and effective field theory.  However, as observed in \cite{Hartnoll:2008vx}, the large-$q$ limit doesn't commute with the small-temperature limit.  So an outstanding question is what the ground state of the system is.

In \cite{Gubser:2008wz}, a ground state was proposed for the choice
 \eqn{VQuartic}{
  V = -{6 \over L^2} + m^2 |\psi|^2 + {u \over 2} |\psi|^4 \,,
 }
with $m^2<0$ and $u>0$.  On the gravity side, this ground state has the structure of a domain wall.  On the ultraviolet side is $AdS_4$ with $\psi=0$.  On the infrared side is $AdS_4$ with $|\psi| = \sqrt{-m^2/u}$.  In between is a charged condensate of $\psi$.  Asymptotically far into the infrared, there is no electric field.  In terms of the dual field theory, this means that all the charge is carried by the condensate: none remains in the normal state.  This is a satisfying picture, but there are some strange features.  The infrared $AdS_4$ signals the emergence of relativistic conformal symmetry, $SO(3,2)$, in the infrared.  One consequence of this symmetry is that there is a maximum speed of propagation for signals at far infrared energies, and this speed is less than the speed of light.  In other words, the geometry has a non-trivial index of refraction for infrared signals.  A second consequence is that the real part of the conductivity exhibits power-law decay for small $\omega$: $\Re\sigma \propto \omega^\delta$ with an exponent $\delta$ that varies continuously as a function of $q$, $m^2$, $u$, and $L$.  Power-law decay at small frequency means that there is no gap.  It was suggested in \cite{Gubser:2008wz} that for a more general class of potentials, a similar domain wall structure would arise, but the infrared side would not be anti-de Sitter.  Instead it would be a warped geometry whose isometry group is the Lorentz group $SO(2,1)$ (together with translation invariance in space and time).  This would again imply a non-trivial index of refraction, but the conductivity would not be forced to exhibit power-law scaling at small $\omega$: it might instead exhibit gapped behavior.

The aim of the current paper is to go beyond the linearized calculations of \cite{Gubser:2008px} and the probe approximation of \cite{Hartnoll:2008vx} and track the thermodynamically preferred phase of the Abelian Higgs model, from $T_c$ down to low temperatures.  This task proceeds largely by finding numerical solutions to the equations of motion, of the form
 \eqn{ansatz}{
  ds^2 &= e^{2A(r)} \left[ -h(r) dt^2 + (dx^1)^2 + (dx^2)^2 \right] + 
    {dr^2 \over h(r)}  \cr
  \Phi &= \Phi(r) \qquad
  \psi = \psi(r) \,,
 }
where $\Phi = A_0$ is the timelike component of the gauge potential.  We always insist that $h(r)$ should have a simple zero at a finite value of $r$: that is, there is always a regular, finite-temperature horizon in our solutions.  In order for $A_\mu dx^\mu = \Phi dt$ to be well-defined at the horizon, we must have $\Phi=0$ at the horizon.  We also insist that $\psi \propto e^{-\Delta_\psi A}$ near the boundary of $AdS_4$, where $\Delta_\psi$ is the larger root of
 \eqn{DeltaPsiDef}{
  \Delta_\psi (\Delta_\psi-3) = m^2 L^2 \,.
 }
This condition is equivalent to specifying that the lagrangian of the dual field theory isn't deformed by ${\cal O}_\psi$.\footnote{For a range of dimensions, $3/2 < \Delta_\psi < 5/2$, it is consistent to choose instead the smaller root of \eno{DeltaPsiDef}.  This choice corresponds to a different CFT, whose correlators can be systematically related to the original CFT \cite{Klebanov:1999tb}.  A discussion of superconducting black holes with these alternative boundary conditions has been given, in the probe approximation, in \cite{Hartnoll:2008vx}.}

We have two main findings:
 \begin{enumerate}
  \item Superconducting black holes of the form \eno{ansatz} are thermodynamically preferred over RNAdS black holes below a temperature $T_c$ at which a second order phase transition occurs.  This was speculated to be the case in \cite{Gubser:2008px} and asserted to be true in the probe approximation in \cite{Hartnoll:2008vx}; also there was a claim in \cite{Gubser:2008zu} of unpublished calculations verifying that superconducting black holes are favored.  More recently, the phase transition was studied in the probe approximation in the more general setting of non-zero superfluid velocity, and a tricritical point was found \cite{Basu:2008st,Herzog:2008he}.  While this paper was nearing completion, we received \cite{Hartnoll:2008kx}, which has some results overlapping ours on thermodynamics away from the probe approximation.  The results reported here together with the literature just cited present quite a convincing case that superconducting black holes describe the preferred phase below the temperature $T_c$ obtained from analyzing linear perturbations around RNAdS.
  \item At least for large enough values of $qL$, an $SO(2,1)$ symmetry and a non-trivial index of refraction arise at sufficiently small temperatures.  This is associated with charge being completely expelled from the black hole, or in field theory terms, to charge being completely carried by the condensate.  As $qL$ decreases, the index of refraction decreases.  For sufficiently small $qL$, it is difficult to say whether $SO(2,1)$ symmetry arises in the limit of small temperatures.  If it does, the relevant temperature scale is very small indeed.
 \end{enumerate}
We have been unable to construct symmetry-breaking solutions at zero temperature.  Such solutions would be dual to the ground state of the conformal field theory in the presence of a finite charge density.  It is reasonable to expect that such solutions exist, and that they exhibit an emergent Lorentz symmetry in the infrared, at least when $qL$ is not too small.

The rest of this paper is organized as follows.  In section~\ref{ANALYTICAL} we collect analytical results on the equations that determine the background as well as equations that can be solved to find the conductivity.  In section~\ref{NUMERICS} we report on the result of our numerical studies of these equations, focusing on thermodynamics, a finite index of refraction, and the conductivity.

\section{Analytical results}
\label{ANALYTICAL}

Plugging the ansatz \eno{ansatz} into the equations of motion resulting from \eno{Lagrangian}, one finds the following second order differential equations:
 \begin{eqnarray}
  A'' &=& -{1 \over 2} \psi'^2 - 
    {q^2 \over 2h^2 e^{2A}} \Phi^2 \psi^2  \label{Aeom} \\
  h'' + 3 A' h' &=& e^{-2A} \Phi'^2 + 
    {2q^2 \over h e^{2A}} \Phi^2 \psi^2  \label{heom} \\
  \Phi'' + A' \Phi' &=& {2q^2 \over h} \Phi \psi^2  \label{Phieom} \\
  \psi'' + \left( 3A' + {h' \over h} \right) \psi' &=& 
    {1 \over h} {\partial V \over \partial\psi^*} - 
      {q^2 \over h^2 e^{2A}} \Phi^2 \psi \,,\quad  \label{psieom}
 \end{eqnarray}
where primes denote $d/dr$.  There is also a first order constraint, which if satisfied at one value of $r$ must hold everywhere, provided the equations of motion (\ref{Aeom}-\ref{psieom}) are also satisfied:
 \eqn{constraint}{
  h^2 \psi'^2 + e^{-2A} q^2 \Phi^2 \psi^2 - 
   {1 \over 2} h e^{-2A} \Phi'^2 - 2 hh' A'  
   - 6 h^2 A'^2 -
   h V(\psi) = 0 \,.
 }
We assume that $\psi$ is everywhere real.  This makes sense because it costs energy for its phase to vary, and with variation of fields only in the $r$ directions, there is no non-trivial topology for its phase to wind around.

Precisely the equations \eno{Aeom}-\eno{constraint} were derived in \cite{Gubser:2008wz}.  But unlike in that paper, here we require that $h$ has a simple zero at a finite value of $r$.  Using the freedom to shift $r$ by an additive constant, we may require that the horizon occurs at $r=0$.  Using also the freedom to rescale $t$ and $\vec{x} = (x^1,x^2)$ by separate multiplicative factors, we may also require
 \eqn{HorizonConstraints}{
  A(0) = 0 \qquad\hbox{and}\qquad h'(0) = 1 \,.
 }
In order to specify a solution to the equations of motion with a regular horizon, we need only two further conditions:
 \eqn{PhiPsiConditions}{
  \psi(0) = \psi_0 \qquad\hbox{and}\qquad \Phi'(0) = \Phi_1 \,.
 }
A standard parameter-counting argument suffices to show that there is now a well-defined Cauchy problem.  Here is how the argument goes.  The four second order equations have eight integration constants, but one is used up by the constraint \eno{constraint}, one more is used up by insisting that the horizon is at $r=0$, and four more are used up by the explicit conditions \eno{HorizonConstraints}-\eno{PhiPsiConditions}.  This appears to leave two constants undetermined.  But in fact, the existence of a regular horizon implicitly requires that $\Phi=0$ at the horizon and that $\psi$ is regular there, and these two conditions amount to fixing the final two integration constants.  So, based on an analysis of the horizon boundary conditions, there is a two-parameter family of solutions, parameterized by $\psi_0$ and $\Phi_1$.

There is one additional boundary condition at infinity: as we have already mentioned, we require that $\psi \propto e^{-\Delta_\psi A}$ near the boundary of $AdS_4$.  This requirement amounts to a non-linear constraint on $(\Phi_1,\psi_0)$.  Any pair of values that satisfies the constraint corresponds to a black hole that spontaneously breaks the abelian gauge symmetry: that is, a superconducting black hole.  In practice, one must make some further restrictions on the allowed class of solutions: for large $r$, $A$ must asymptote to a linear function of $r$ with positive slope, while $h$ and $\Phi$ must asymptote to constants.  There are solutions which violate one or more of these restrictions: for example, $A \to -\infty$ as $h \to +\infty$ is fairly common.  We believe all such solutions are singular.  Certainly they are not asymptotically $AdS_4$, and as such they can naturally be excluded.  By studying series expansions near the boundary of solutions that are asymptotically $AdS_4$, one can show that 
 \eqn{LargeR}{
  A(r) &= a_1 r + a_0 + \ldots  \cr
  h(r) &= H_0 + H_3 e^{-3A} + \ldots \cr
  \Phi(r) &= p_0 + p_1 e^{-A} + \ldots  \cr
  \psi(r) &= \Psi_0 e^{-\Delta_\psi A} + \ldots \,,
 }
where in each equation, $\ldots$ indicates terms which are subleading at large $r$ to the ones shown.  Because $r=0$ is by assumption the horizon, i.e. the largest value of $r$ where $h$ has a zero, we must have $H_0>0$; also, as remarked previously, we require $a_1>0$; but all the other constants in \eno{LargeR} might in principle have either sign.  In practice, we generally find $p_0>0$ and $p_1<0$.

We make one further restriction on the solutions we study below: $\psi(r)$ is not allowed to have any nodes.  This is based on the idea that oscillations in $\psi$ can only add to the energy, and there is no topology to support them, so a solution with nodes is probably unstable toward decay to a solution with no nodes.  Such expectations have been borne out in related studies of solutions of the Einstein-Yang-Mills equations: for a recent review see \cite{Winstanley:2008ac}.

Having found a solution to \eno{Aeom}-\eno{constraint}, one can determine its thermodynamics by extracting the coefficients shown in \eno{LargeR}.  The energy density, entropy density, temperature, chemical potential, and charge density can be read off as
 \eqn{ThermoFormulas}{\seqalign{\span\TL & \span\TR & \qquad\qquad \span\TL & \span\TR}{
  \epsilon &= -{H_3 \over \kappa^2 L H_0} &
  s &= {2\pi \over \kappa^2} e^{2A(0)}  \cr
  \mu &= {p_0 \over 2L\sqrt{H_0}} &
  T &= {1 \over 4\pi} e^{A(0)} {h'(0) \over \sqrt{H_0}}  \cr
  \rho &= -{p_1 \over \kappa^2 \sqrt{H_0}} &
  f &= {\epsilon - Ts} \,.
 }}
We adhere to conventions of \cite{Gubser:2008zu}, which discusses the thermodynamics of a model similar to the Abelian Higgs in $AdS_4$.  Energy density, temperature, and chemical potential are measured with respect to the Killing time $\sqrt{H_0}t$ instead of $t$.  We also follow \cite{Gubser:2008zu} in defining rescaled thermodynamic quantities with no $\kappa$ dependence, as follows:
\eqn{RescaledThermo}{\seqalign{\span\TL & \span\TR & \qquad\qquad \span\TL & \span\TR}{
  \hat\epsilon &= {\kappa^2 \epsilon \over (2\pi)^3 L^2} &
 \hat{s} &= {\kappa^2 s \over (2\pi)^3 L^2} \cr
  \hat\rho &= {\kappa^2 \rho \over (2\pi)^3 L^2} &
 \hat{f} &= {\kappa^2 f \over (2\pi)^3 L^2}  \,.
 }}
The states in the dual conformal field theory that charged black holes describe are characterized by two energy scales, which we can choose to be $\sqrt{\hat\rho}$ and $T$.  The dimensionless ratio $T/\sqrt{\hat\rho}$ parameterizes a family of symmetry-breaking black hole solutions in the $\Phi_1$-$\psi_0$ plane.  Let us consider $\hat\rho$ to be fixed.  At some critical temperature, $T=T_c$, the family of solutions intersects the flat line at $\psi_0=0$ that corresponds to RNAdS.  It is sometimes convenient to use $T/T_c$ instead of $T/\sqrt{\hat\rho}$ to parameterize the broken solutions.  In doing so, we should note that we hold $\hat\rho$ fixed: that is, what $T/T_c$ really means is $(T/\sqrt{\hat\rho}) / (T_c/\sqrt{\hat\rho_c})$.  This is distinct from what we would get by holding the chemical potential $\mu$ fixed.  Physically, we are studying the boundary theory at fixed charge density.

Following \cite{Hartnoll:2008vx,Gubser:2008wz}, we will also explore the conductivity for a selection of the solutions we generate.  Because the condensate is purely $s$-wave, the conductivity is the same in any direction: it is a complex scalar, $\sigma$, whose real part measures the dissipative response and whose imaginary part measures the reactive response.  One may calculate the conductivity using the formula $\sigma(\omega) = J(\omega)/E(\omega)$, where $J$ is the current response to a spatially homogeneous electric field $E$.  More precisely, if we introduce a perturbed gauge field,
 \eqn{PerturbedA}{
  A_\mu dx^\mu = \Phi dt + e^{-i\omega\sqrt{H_0} t} a_x(r) dx \,,
 }
and also perturb the metric by introducing an off-diagonal element
 \eqn{PerturbedMetric}{
  g_{tx} = e^{-i\omega\sqrt{H_0} t} e^{2A(r)} h_{tx}(r) \,,
 }
then the linearized equations of motion couple $a_x$ and $h_{tx}$ but do not require any additional fields to be perturbed.  Allowing complex perturbations to the gauge field and the metric is a formal trick: it is understood that we take the real part at the end of the day.  The linearized Einstein equations reduce to a single constraint between $h_{tx}$ and $a_x$:
 \eqn{htxpSolve}{
  h_{tx}' + e^{-2A} \Phi' a_x = 0 \,.
 }
This constraint can be used to eliminate $h_{tx}$ from the linearized Maxwell equations.  The resulting equation for $A_x$ is
 \eqn{axEom}{
  a_x'' + \left( A' + {h' \over h} \right) a_x' +
   {1 \over h} \left( {\omega^2 H_0 \over h e^{2A}} - 2q^2 \psi^2 - 
      {\Phi'^2 \over e^{2A}} \right) a_x = 0 \,.
 }
Essentially this equation also appeared in \cite{Gubser:2008wz}.  The perturbation $a_x$ must obey standard horizon boundary conditions: it is infalling.  With the requirements \eno{HorizonConstraints}, this simply means that $a_x \propto r^{-i\omega\sqrt{H_0}}$ for small $r$.  At large $r$, a series solution in powers of $e^{-r}$ suffices to show that
 \eqn{axFar}{
  a_x(r) = a_x^{(0)} + a_x^{(1)} e^{-A(r)} + \ldots  \,.
 }
The conductivity is
 \eqn{Conductivity}{
  \sigma &= -{i \over \omega} {a_x^{(1)} \over a_x^{(0)}} \,,
 }
simply because $a_x^{(1)} e^{-i\omega\sqrt{H_0} t}$ is the expectation value of the current $J_x$ in the dual field theory, while $i\omega a_x^{(0)} e^{-i\omega\sqrt{H_0} t}$ is the electric field.

Before delving into the results of numerically solving \eno{Aeom}-\eno{constraint}, let us note an argument \cite{Gubser:2005ih} in favor of the existence of symmetry-breaking black holes, for sufficiently large $qL$, that avoids numerics entirely and employs instead the Breitenlohner-Freedman (BF) bound \cite{Breitenlohner:1982bm,Breitenlohner:1982jf}.\footnote{We note that a related line of argument has been refined and extended in \cite{Hartnoll:2008kx} to the case of neutral scalars with $m^2 L^2 = -2$.}  The starting point is to note that the $T \to 0$ limit of the Reissner-Nordstrom black hole \eno{RNAdS} has for its near-horizon geometry $AdS_2 \times {\bf R}^2$.  Let's choose the radial variable $r$ in \eno{RNAdS} so that the horizon is at $r_H=0$.  Then at extremality one has
 \eqn{ERNAdS}{
  \hat\epsilon = {4 \over (2\pi)^3 L^3} \qquad
   \hat\rho = {\sqrt{12} \over (2\pi)^3 L^2} \,,
 }
where $\hat\epsilon$ and $\hat\rho$ are defined as in \eno{RescaledThermo}.  One easily sees that the radius of the $AdS_2$ geometry near $r=0$ is
 \eqn{LIR}{
  L_{\rm IR} = {L \over \sqrt{6}} \,.
 }
The transverse space is not significant for what follows (in the context of \cite{Gubser:2005ih} it was $S^2$ not ${\bf R}^2$).  The point is to note that the BF bound points to an instability in $AdS_2$ when a scalar has a effective mass $m_{\rm IR}^2$ satisfying 
 \eqn{ViolateBF}{
  m_{\rm IR}^2 L_{\rm IR}^2 < -{1 \over 4} \,.
 }
Here $m_{\rm IR}^2$ is the limit as $r \to r_H$ of the effective mass squared defined in \cite{Gubser:2008px}:
 \eqn{meff}{
  m_{\rm eff}^2 = m^2 + g^{tt} q^2 \Phi^2 \,.
 }
When the inequality \eno{ViolateBF} holds, according to the logic of \cite{Gubser:2005ih}, one should expect not just a few branches of symmetry-breaking solutions, but infinitely many, with arbitrarily many nodes in $\psi$.  If the inequality does not hold, it is not clear whether or not superconducting solutions exist: the question then depends on more detailed properties of the matter lagrangian.  But at most one expects only finitely many branches of symmetry-breaking solutions.  As usual, the branches with nodes can generally be expected to be unstable and/or thermodynamically disfavored.

A short calculation based on \eno{RNAdS} and \eno{meff} shows that \eno{ViolateBF} is equivalent to
 \eqn{BFequivalent}{
  m^2 L^2 - 2 q^2 L^2 < -{3 \over 2} \,.
 }
For comparison, the BF bound in $AdS_4$ is $m^2 L^2 \geq -9/4$.  If $m^2 L^2 < -9/4$, then empty $AdS_4$ itself is unstable toward developing non-zero $\psi$.  If $-9/4 < m^2 L^2 < -3/2$, then $AdS_4$ is stable, but the argument explained in the previous paragraph indicates that there should be symmetry-breaking charged black hole solutions no matter what value $q$ takes.  If $m^2 L^2 > -3/2$, then only for sufficiently large $q$ does this argument imply the existence of symmetry-breaking charged black hole solutions.  Our numerics will focus on $m^2 L^2 = -2$ and $qL$ between $0.1$ and $2$.\footnote{We thank S.~Hartnoll for correcting a numerical error in a previous version of this discussion.}

\section{Numerical results}
\label{NUMERICS}

As explained following \eno{PhiPsiConditions}, the black hole solutions we are interested in can be parameterized by $\psi_0$ (the value of the scalar field at the horizon) and $\Phi_1$ (roughly, the electric field in the $r$ direction at the horizon).  The boundary condition on the scalar at the conformal boundary of anti-de Sitter space amounts to a single non-linear constraint between $\psi_0$ and $\Phi_1$.  There is thus a one-parameter family of solutions.  More precisely, there are several such families, or perhaps even infinitely many, with each family corresponding to black hole solutions where $\psi$ has zero, one, two, or more nodes.  As explained following \eno{LargeR}, we focus our attention on the family of solutions where $\psi$ is everywhere positive.  To explore this family of solutions, we started by finding a few symmetry-breaking solutions with small $\psi_0$.  These solutions sufficed to determine $T_c$ with good accuracy.  ($T_c$ can also be determined by treating the scalar as a linearized perturbation of the Reissner-Nordstrom-anti-de Sitter solution, as in \cite{Gubser:2008px}.)  Then we proceeded to larger $\psi_0$ in small steps, ensuring at each step that the boundary condition on the scalar is satisfied.  We iterated this process until we ran out of CPU time (the allocation was roughly $12$ hours per branch) or until we encountered numerical errors suggesting that the solutions were no longer reliable.

We further restrict our attention to the scalar potential \eno{VQuadratic} with $m^2 L^2 = -2$.  As explained around \eno{DeltaPsiDef}, this corresponds to an operator ${\cal O}_\psi$ in the dual field theory with dimension $2$, provided one makes the usual assignment of operator dimensions.  There are three dimensionful parameters in the lagrangian: $\kappa$, $L$, and $q$.  But $\kappa$ enters only as an overall scale which doesn't affect the classical equations of motion or rescaled thermodynamic quantities.  Only the dimensionless combination $qL$ can enter into formulas describing the physics of the dual CFT, heuristically because conformal invariance prohibits any dimensionful scale in the theory.

\subsection{Thermodynamics}

 \begin{figure}
  \centerline{\includegraphics[width=6in]{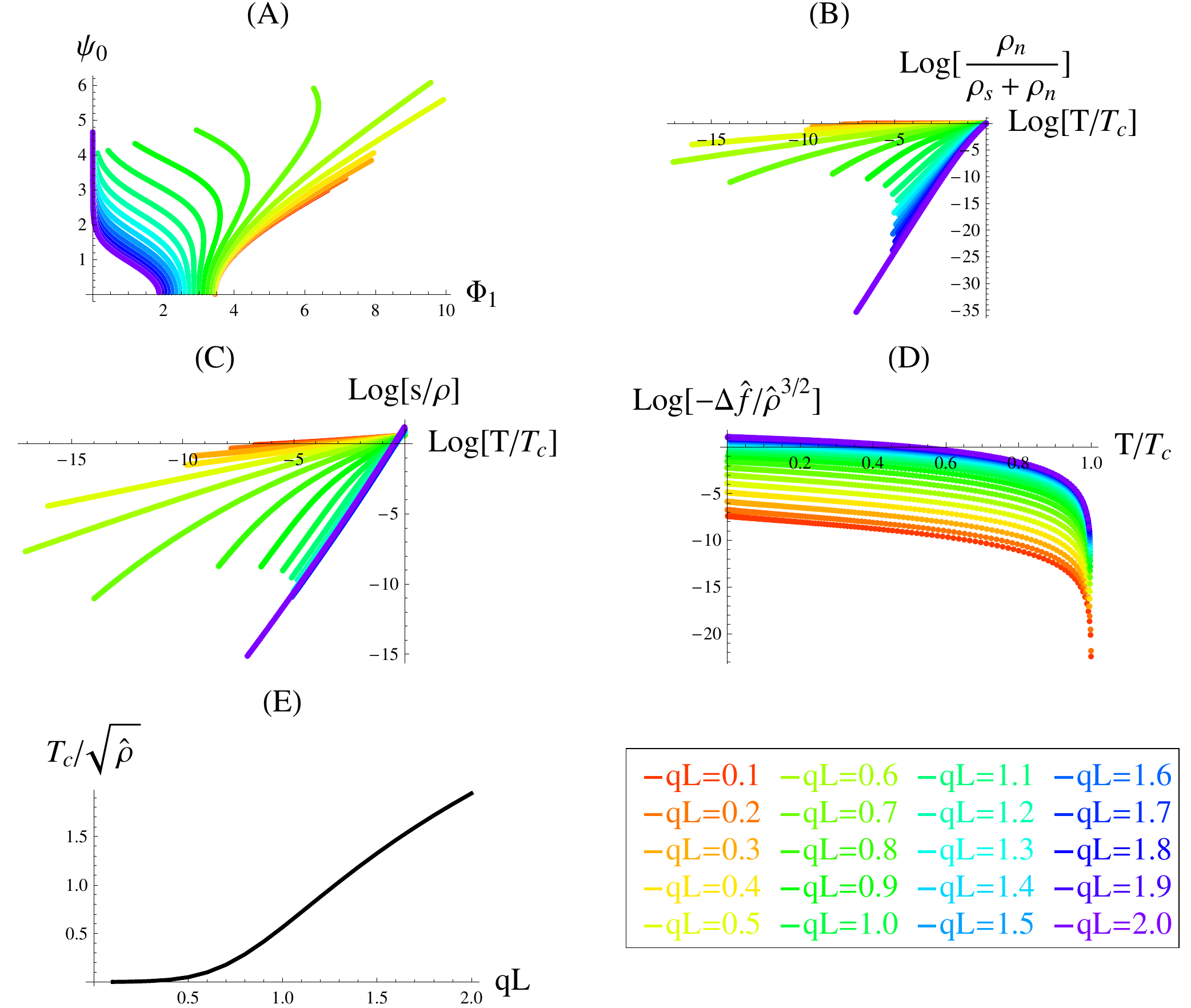}}
  \caption{(Color online) Thermodynamic quantities as functions of $qL$ and temperature.  In (A)-(D), the curves for each value of $qL$ have different lengths because they terminate either where numerical errors began to cast doubt on the validity of the coldest solutions, or where we simply ran out of CPU time.  All logarithms are natural logs.}\label{SCALINGPLOTS}
 \end{figure}
Figure~\ref{SCALINGPLOTS} shows a summary of thermodynamic quantities as a function of $qL$ and the temperature.
Of particular interest is $\Delta\hat{f}$, which is the rescaled free energy of the superconducting black hole, minus the rescaled free energy of the Reissner-Nordstrom black hole with the same temperature and charge density.  The criterion for superconducting black holes to be preferred in the microcanonical ensemble is $\Delta\hat{f} < 0$.  As is evident from figure~\ref{SCALINGPLOTS}D, $\Delta\hat{f}$ is indeed negative for $T<T_c$, no matter what $qL$ is, over the range $0.1 < qL < 2$.  It appears that there are thermodynamically preferred superconducting black holes for all values of $qL$ when the potential is \eno{VQuadratic} with $m^2 L^2 = -2$.  This agrees with our BF bound argument from section~\ref{ANALYTICAL}.  If we chose $m^2 L^2$ less negative, or positive, then we would expect there to be a minimum value of $qL$ below which superconducting black holes either do not exist or are not thermodynamically favored.  But, according to the arguments of \cite{Gubser:2008px}, we expect that sufficiently large $qL$ should produce superconducting black holes for any fixed value of $m^2 L^2$.

\subsection{Index of refraction}

\begin{figure}
  \centerline{\includegraphics[width=6in]{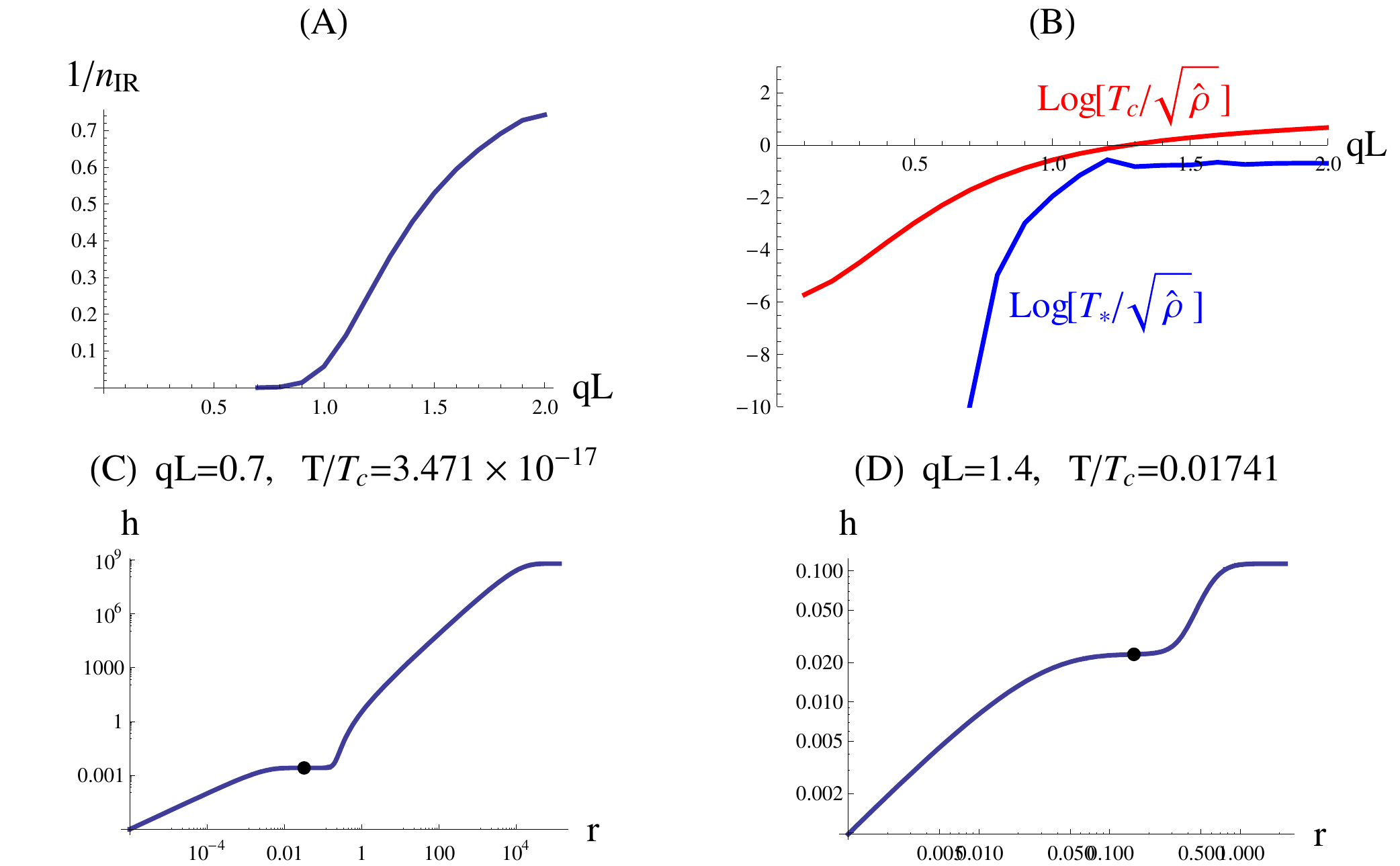}}
  \caption{(Color online) (A) The inverse index of refraction, $1/n_{\rm IR}$, as a function of $qL$.  (B) The critical temperature $T_c$ and the inflection point temperature $T_*$ compared to the rescaled charge density $\hat\rho$ as functions of $qL$.  (C) For $qL=0.7$ and the coldest temperature we could reach, there is an inflection point in $h(r)$ at the location of the black dot.  In order to see a distinctive double shelf develop around this inflection point, we had to cool the black hole quite a bit more than we did for figure~\ref{SCALINGPLOTS}.  (D) For $qL=1.4$, there is a distinctive double-shelf structure in $h(r)$ at a less extreme temperature than for $qL=0.7$.  The black dot is again at an inflection point of $h(r)$.}\label{NIRPLOTS}
 \end{figure}

At least for $qL \gsim 0.7$, a distinctive feature emerges at low temperatures: $h$ has a ``double-shelf'' structure, visible in figures~\ref{NIRPLOTS}C and~\ref{NIRPLOTS}D.  If the lower shelf extended infinitely far into the infrared, instead of terminating in a regular horizon, then we would wind up with a domain wall geometry similar to the one exhibited explicitly in \cite{Gubser:2008wz}.  In such a case, the infrared geometry would not be anti-de Sitter space, but rather some more general warped geometry with $SO(2,1)$ invariance.  The overall geometry would then be a domain wall with no boost invariance separating two asymptotic geometries, both of which have $SO(2,1)$ invariance, but with different values of $h$, call them $h_{\rm UV}$ and $h_{\rm IR}$.  The overall normalization of $h$ can be changed by reparameterizing $r$, but the ratio
 \eqn{nIRdef}{
  n_{\rm IR} \equiv \sqrt{h_{\rm UV} \over h_{\rm IR}}
 }
is an invariant quantity.  It can be termed an index of refraction, because it characterizes the maximum speed of transmission of signals in the infrared.
Unlike in \cite{Gubser:2008wz}, we have been unable to explicitly construct zero-temperature solutions with an infinite shelf.  However, examination of a series of solutions with the same $qL$ and decreasing values of $T/T_c$ shows that the lower shelf broadens out without changing its height appreciably.  Because the existence of a shelf is a somewhat qualitative criterion, we studied instead the inflection points of $h$ as a function of $r$.  There are never more than two.  If there are two, then the one at the smallest value of $r$ is probably associated with double shelf behavior.  We define $n_{\rm IR}$ for a finite-temperature solution by replacing $h_{\rm IR}$ by $h$ evaluated at its first inflection point.  In figure~\ref{NIRPLOTS}A, we show $1/n_{\rm IR}$ for the coldest black hole we could construct as a function of $qL$.
For $qL \lsim 0.7$, we couldn't find solutions where there are two inflection points.  For larger values, we observe that there is a temperature $T_*$ below which there are two inflection points, and above which there are not.  We are not sure whether the double shelf appears at all values of $qL$ or not.  If it does, $T_*$ quickly becomes small, and $n_{\rm IR}$ quickly becomes large, as $qL$ decreases.  Indeed, we found $n_{\rm IR} \approx 4.8 \times 10^{5}$ at $qL = 0.7$.

\subsection{Conductivity}
\begin{figure}
  \centerline{\includegraphics[width=6in]{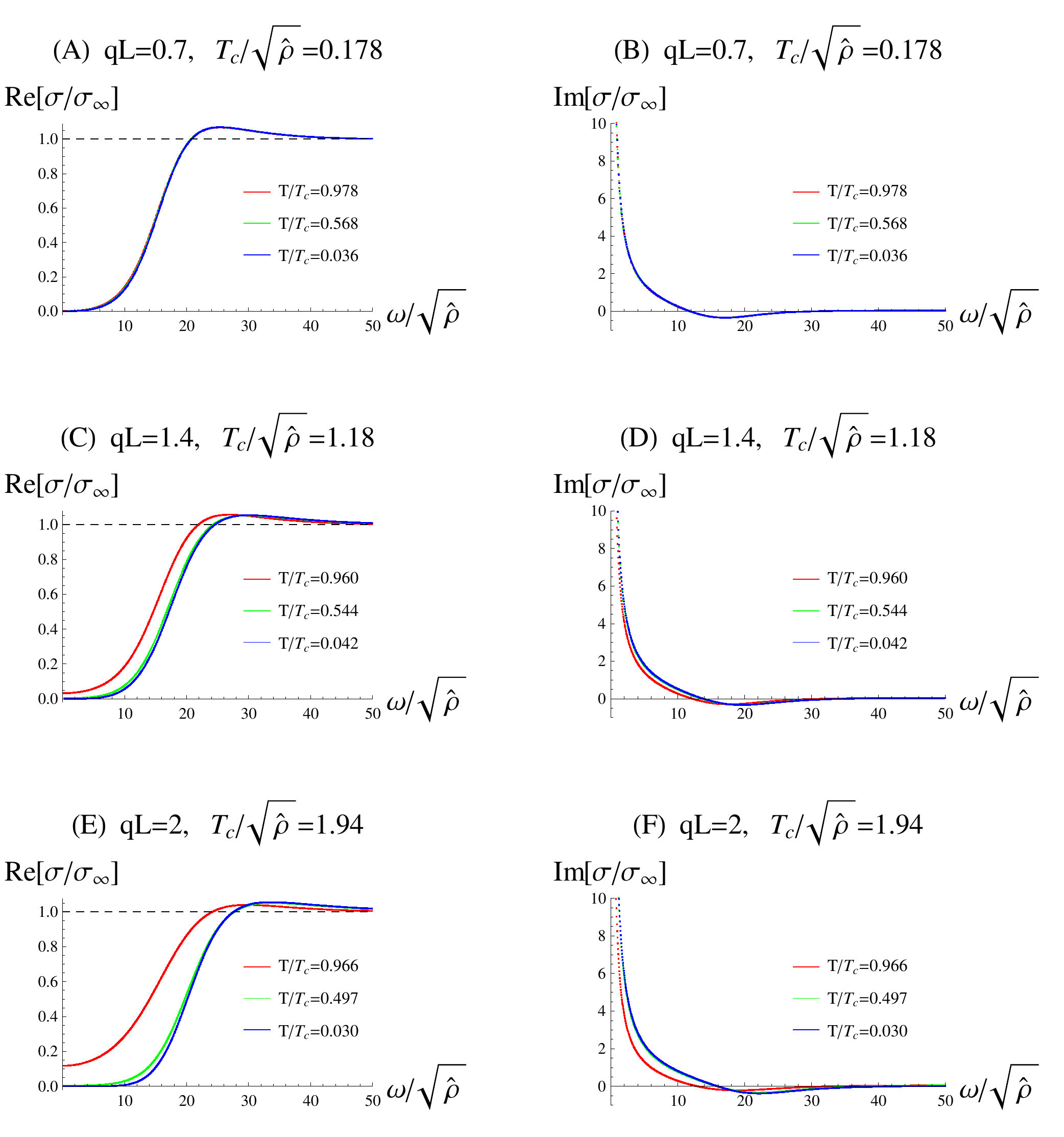}}
  \caption{(Color online) (A), (B) The real and imaginary parts of the conductivity for $qL=0.7$ at various values of $T/T_c$.  (C), (D) The real and imaginary parts of the conductivity for $qL=1.4$ at various values of $T/T_c$.  (E), (F) The real and imaginary parts of the conductivity for $qL=2$ at various values of $T/T_c$.}\label{CONDUCTIVITYPLOTS}
 \end{figure}

For a selection of superconducting black holes, we computed the conductivity as a function of frequency by solving \eno{axEom} with infalling boundary conditions at the horizon and using \eno{Conductivity}.  Figure~\ref{CONDUCTIVITYPLOTS} shows the real and imaginary parts of the conductivity at various temperatures for $qL=0.7$, $1.4$, and $2$.
The results are roughly in line with the earlier results \cite{Hartnoll:2008vx} in the probe approximation; also, they appear to approximately agree with the results of \cite{Hartnoll:2008kx}, though the range of $qL$ we surveyed is further from the probe approximation.

Several qualitative points are worth noting:
 \begin{itemize}
  \item The behavior $\Im\sigma \propto 1/\omega$ indicates the presence of a $\delta(\omega)$ contribution to $\Re\sigma$.
  \item When $qL=0.7$, the conductivity curves appear to lie on top of each other.  One way to understand this is that $T_c$ is very small compared to $\sqrt{\hat\rho}$ for $qL = 0.7$, and making $T$ even smaller doesn't significantly change the way the system responds.
  \item The behavior of $\sigma$ at small $\omega/\sqrt{\hat\rho}$ is more ``gap-like'' for larger values of $qL$: that is, $\sigma$ is closer to $0$ for a wider interval.  But we do not find evidence for $\sigma$ being strictly zero up to some finite $\omega$.  It could be that this happens for $T$ exactly equal to $0$; but we are inclined to think that there isn't a ``true gap'' in these systems.
 \end{itemize}
It is notable that a finite $\delta(\omega)$ appears in $\Re\sigma$ even as $T$ approaches $T_c$---in fact, even above $T_c$ it persists.  As noted in \cite{Hartnoll:2008kx}, this is associated with the translation invariance of the system, which prevents a DC current from relaxing.  Thus, infinite DC conductivity is {\it not} a feature that distinguishes the superconducting black holes from the normal state RNAdS black holes; rather, the breaking of the abelian gauge symmetry through the formation of the condensate is the distinguishing feature.

\section*{Acknowledgments}

We thank C.~Herzog, F.~Rocha, and S.~Pufu for useful discussions.  This work was supported in part by the Department of Energy under Grant No.\ DE-FG02-91ER40671 and by the NSF under award number PHY-0652782. 

\bibliographystyle{ssg}
\bibliography{photon}
\end{document}